\documentclass[prc,preprint]{revtex4}

\usepackage{graphicx}

\begin{document}

\title{Asymmetry energy of nuclear matter:
Temperature and density dependence, and validity of semi-empirical formula}
\author{\bf G. H. Bordbar, R. Feridoonnezhad and M. Taghizade}
 \affiliation{Physics Department, Shiraz University, Shiraz 71454, Iran\\
and\\
Center for Excellence in Astronomy and Astrophysics (CEAA-RIAAM)-Maragha, P.O. Box 55134-441, Maragha 55177-36698, Iran}
\begin{abstract}
In this work, we have done a completely microscopic calculation using a many-body variational method based on the cluster
expansion of energy to compute the asymmetry energy of nuclear matter. In our calculations, we have employed the
$AV_{18}$ nuclear potential. We have
also investigated the temperature and density dependence of asymmetry energy. Our results show
that the asymmetry energy of nuclear matter depends on both density and temperature. We have also
studied the effects of different terms in the asymmetry energy of nuclear matter. These
investigations indicate that at different densities and temperatures, the contribution of parabolic
term is very substantial with respect to the other terms. Therefore, we can conclude that the
parabolic approximation is a relatively good estimation, and our calculated binding energy of
asymmetric nuclear matter is in a relatively good agreement with that of semi-empirical mass
formula. However, for the accurate calculations, it is better to consider the effects of other terms.
\end{abstract}
\maketitle
\section{INTRODUCTION}
Nuclear matter is an extremely large hypothetical system of constrained nucleons which interact
through the strong nuclear force, and can be considered as an ideal model for the nuclear matter
within the heavy nuclei\cite{1}.
The analysis of heavy nucleus behavior in high density and finite temperature is a challenging issue in modern nuclear physics \cite{2}.
The nuclear matter asymmetry energy is an important subject in nuclear physics and astrophysics. According to research conducted \cite {3,4,5,6,7,8,9,10}, the nuclear matter asymmetry energy plays a crucial role in determining many important nuclear properties such as the neutron skin of nuclear system, structure of nuclei near the drip line, and neutron star mass and radius. They are especial important  constraints on the nuclear matter equation of state and the density dependence of the asymmetry energy \cite{9,11,12}. The nuclear matter equation of state plays a central roles in the study of high-energy in nuclear physics, especially in heavy ion collisions. The nuclear matter equation of state at zero temperature governs the structure of cold neutron stars, the equation of state for finite temperatures is necessary for studies of core collapse supernovae, black hole formation and cooling of the neutron stars \cite{13,14,15}.

Due to the strong interaction between the nucleons, we must completely include the nucleon-nucleon potential in the nuclear matter calculations.
The nuclear potential is usually made out by fitting the nucleon-nucleon scattering data and deuteron properties. The potential models which obtained only from fitting the $np$ scattering data frequently offer a weak description of $pp$ and $nn$ data \cite{16}. Recently, a nuclear potential has been introduced by Wiringa \textit{et al.} which extracted by fitting $np$ and $pp$ data as well as low energy $nn$ scattering parameters and deuteron properties \cite{18}.
This potential is written in an operator format which depends on the value of $J,\
L,\ S$ and $ T$ as well as $T_z$ of nucleon-nucleon pair. Therefore, in the calculations with this potential,
a completely microscopic calculation in which the third
component of isospin ($T_z$) would be explicitly considered is required. This is a very important point in the
calculations for asymmetric nuclear matter consisting of $Z$ protons and $N$ neutrons, where in
general $Z\neq N$. The microscopic studies on asymmetric nuclear matter are very limited, and usually
the asymmetry energy of nuclear matter  is computed employing a
parabolic approximation. Therefore, it is interesting to study the validity of this estimation for the
asymmetric nuclear matter.

One of the powerful microscopic methods in the nuclear matter calculations is the variational method based on the cluster expansion of energy.
In recent years, we have done several works on the nucleonic matter using this method \cite{19,20,21,22,23,24,25,26}.
The purpose of present work is to compute the asymmetry energy of nuclear matter using $AV_{18}$ potential at finite temperature employing the mentioned microscopic technique.  We also verify the validity of semi-empirical mass formula at different temperatures and densities. The dependence of asymmetry energy of asymmetric nuclear matter on density and temperature is also investigated.

\section{Formalism}
Here, a system with $A$ interacting nucleons (including $N$ neutrons and $Z$ protons) is considered in order to calculate the binding energy of asymmetric nuclear matter with $AV_{18}$ potential. As it was mentioned in the preceding  section, we use the variational method based on the cluster expansion of the energy. In this method, at first we choose a trial wave function as follows,
\begin{equation}
\psi=F\Phi,
\end{equation}
where $\Phi$ is the antisymmetric wave function of $A$ non-interacting nucleons and $F=F(1\ldots{}A)$ is the inter-nucleon correlation function included by the interaction between the nucleons. We use the Jastrow ansatz for the correlation function,
\begin{equation}
F=S\prod_{i>j}f(ij),
\end{equation}
where $S$ is the symmetrizing operator and $f(ij)$ is the two-body correlation function.
Using the above trial wave function included by this correlation function, we get a
cluster expansion for the energy which up to the two-body term is as follows,
\begin{equation}
E([f])=\frac{1}{A}\frac{<\psi|H|\psi>}{<\psi|\psi>}=E_1+E_2.
\end{equation}

At finite temperature, the one-body term $E_1$ for an asymmetric nuclear matter consisting of $Z$ protons and $N$ neutrons is as follows,
\begin{equation}\label{E1}
E_1=E_1^{(p)} + E_1^{(n)}.
\end{equation}
Labels $p$ and $n$ are used for the protons and neutrons, and $E_1^{(i)}$ is,
\begin{equation}\label{E1i}
E_1^{(i)}=\sum_{k}\frac{\hbar^2k^2}{2m^{(i)}}n^{(i)}(k,T,\rho^{(i)}),
\end{equation}
where $n^{(i)}(k,T,\rho^{(i)})$ is fermi-Dirac distribution function,
 \begin{equation}
 n^{(i)}(k,T,\rho^{(i)})=\frac{1}{Exp\{\beta[\varepsilon^{(i)}(k)-\mu^{(i)}(T,\rho^{(i)})]\}+1}.
 \end{equation}
In the above equation,  $\mu^{(i)}$ is the chemical potential, $\rho^{(i)}$ is the number density
and $\varepsilon^{(i)}=\frac{\hbar^2k^2}{2m^{(i)}}$ is the single particle energy of nucleon $i$, and $\beta=\frac{1}{k_BT}$ ($T$ is the temperature).
The total density of  system is,
\begin{equation}
\rho=\rho^{(p)}+\rho^{(n)}.
\end{equation}
For the asymmetric nuclear matter, we can also define the asymmetry parameter as follows,
\begin{equation}
\xi=\frac{N-Z}{N+Z}=\frac{\rho^{(n)}-\rho^{(p)}}{\rho^{(n)}+\rho^{(p)}}.
\end{equation}
This definition leads to the following relations for the nucleon density,
\begin{equation}\label{asy}
\rho^{(p)}=(1-\xi)\rho, \ \ \ \ \ \ \ \ \rho^{(n)}=(1+\xi)\rho.
\end{equation}
By comparing Eqs. (\ref{asy}), (\ref{E1i}) and (\ref{E1}), we can conclude that the energy of asymmetric nuclear
matter is different for various asymmetry parameter.

The two-body energy has the following form,
\begin{equation}
E_2=\frac{1}{2A}\sum_{ij}<ij\mid\upsilon(12)\mid ij-ji>,
\end{equation}
where
\begin{equation}
\upsilon(12)=-\frac{\hbar^2}{2m}[f(12),[{\nabla_{12}}^2,f(12)]]+f(12)V(12)f(12).
\end{equation}
$V(12)$ is the nucleon-nucleon pair potential.
The two-body correlation function, $f(12)$, is considered as follows,
\begin{equation}
f(12)=\sum_{p=1}^3f^{(p)}(r_{12})O^{(p)}(12),
\label{f12}
\end{equation}
where operators $O^{(p)}(12)$ are,
\begin{eqnarray}\label{Pcor}
O^{(p=1,3)}(12)&=&(\frac{1}{4}-\frac{1}{4}\sigma_{1}.\sigma_{2}),\
(\frac{1}{2}+\frac{1}{6}\sigma_{1}.\sigma_{2}+\frac{1}{6}S_{12}),\nonumber\\
&&(\frac{1}{4}+\frac{1}{12}\sigma_{1}.\sigma_{2}-\frac{1}{6}S_{12}).
\end{eqnarray}
In above equation, $S_{12}=3(\sigma_1.\hat{r})(\sigma_2.\hat{r})-\sigma_1.\sigma_2$ is the tensor operator.
We employ the $AV_{18}$ nuclear potential in our calculations. This potential has the following form \cite{18},
\begin{equation}
V(12)=\sum_{p=1}^{18}V^{(p)}(r_{12})O_{12}^{(p)}.
\label{V12}
\end{equation}
The first fourteen operators are,
\begin{eqnarray}
O_{12}^{p=1-14}&=&1,\sigma_1.\sigma_2,\tau_1.\tau_2,(\sigma_1.\sigma_2)(\tau_1.\tau_2),S_{12},S_{12}(\tau_1.\tau_2),
\textbf{L}.\textbf{S},\nonumber\\
&&\textbf{L}.\textbf{S}(\tau_1.\tau_2),\textbf{L}^2,\textbf{L}^2(\sigma_1.\sigma_2),\textbf{L}^2(\tau_1.\tau_2),
\textbf{L}^2(\sigma_1.\sigma_2)(\tau_1.\tau_2),\nonumber\\
&&
(\textbf{L}.\textbf{S})^2,(\textbf{L}.\textbf{S})^2(\tau_1.\tau_2),
\end{eqnarray}
and four additional operators which break the charge independence of nuclear force are,
\begin{equation}
O_{12}^{p=15-18}=\textbf{T}_{12},\textbf{T}_{12}(\sigma_1.\sigma_2),S_{12}\textbf{T}_{12},(\tau_{1z}+\tau_{2z}).
\end{equation}
 $\textbf{T}_{12}=3(\tau_1.\hat{r})(\tau_2.\hat{r})-\tau_1.\tau_2$ is the isotensor operator.

Using the above formalism, we can compute the binding energy of asymmetric nuclear matter.
The procedure of these calculations has been fully presented in our
previous articles \cite{19,20,21,22,23,24,25,26}.

\section{Results and Discussions}

\begin{figure}
\centering
\includegraphics[height=0.4\textwidth]{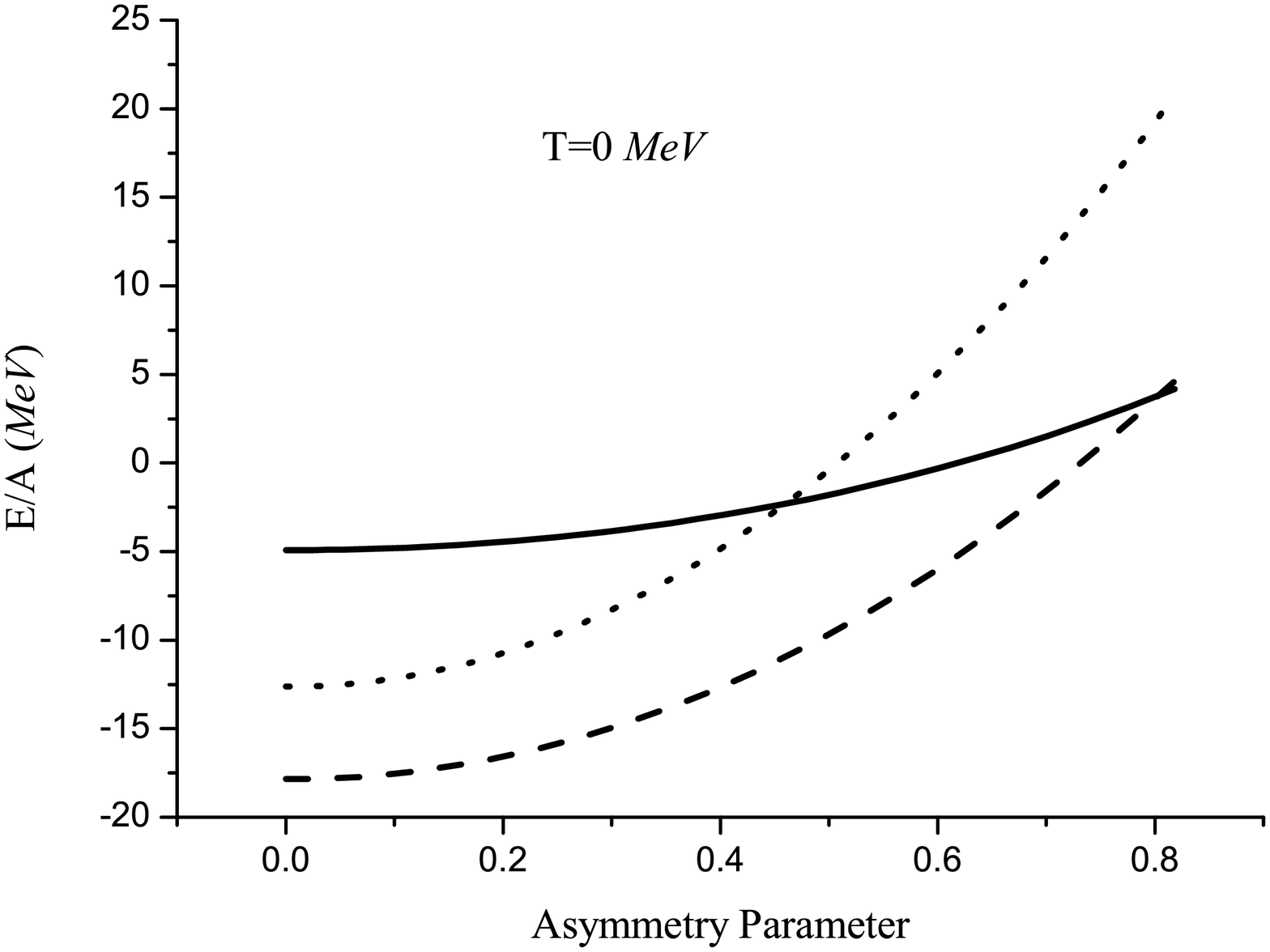}
\includegraphics[height=0.4\textwidth]{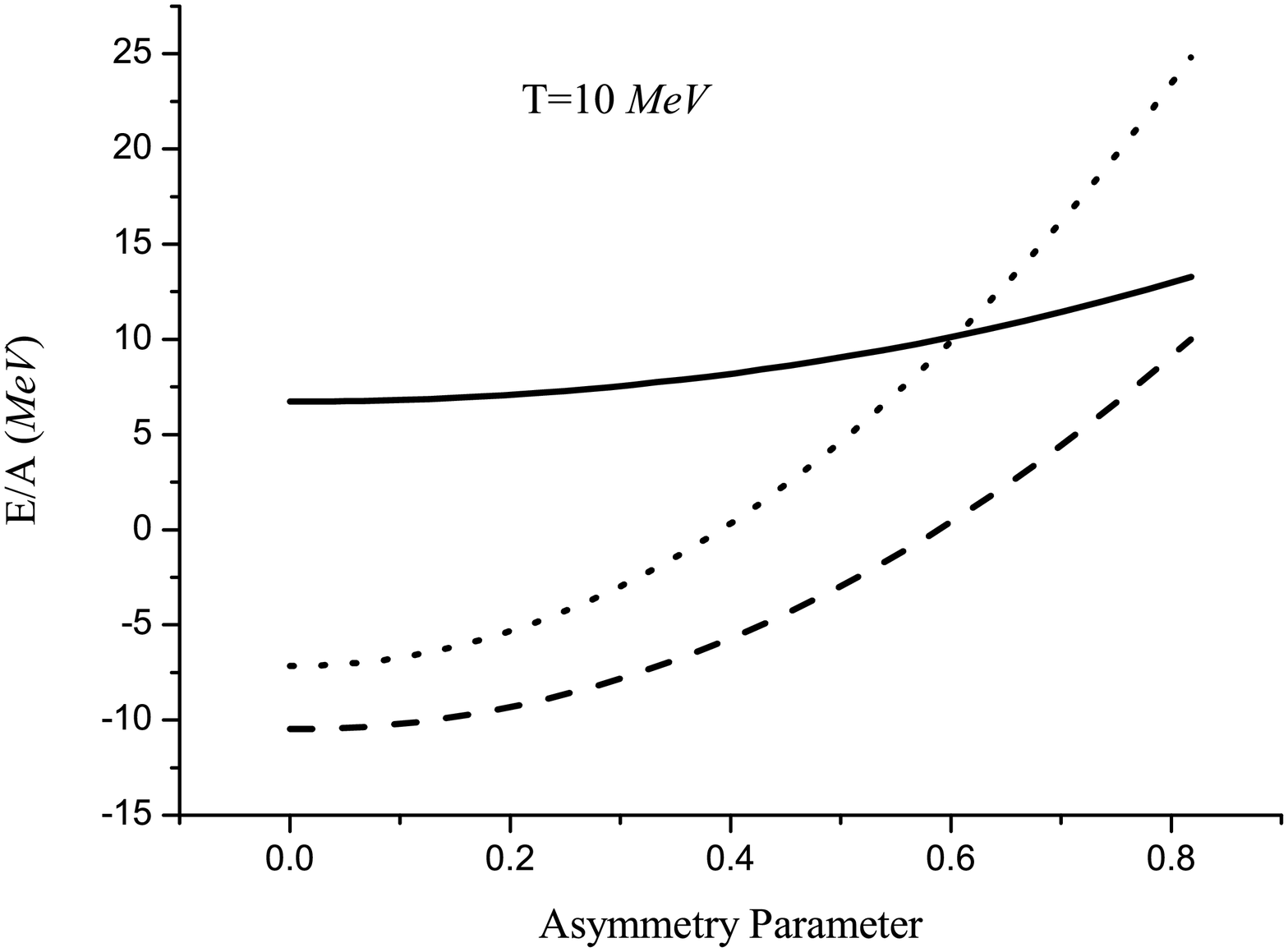}
\includegraphics[height=0.4\textwidth]{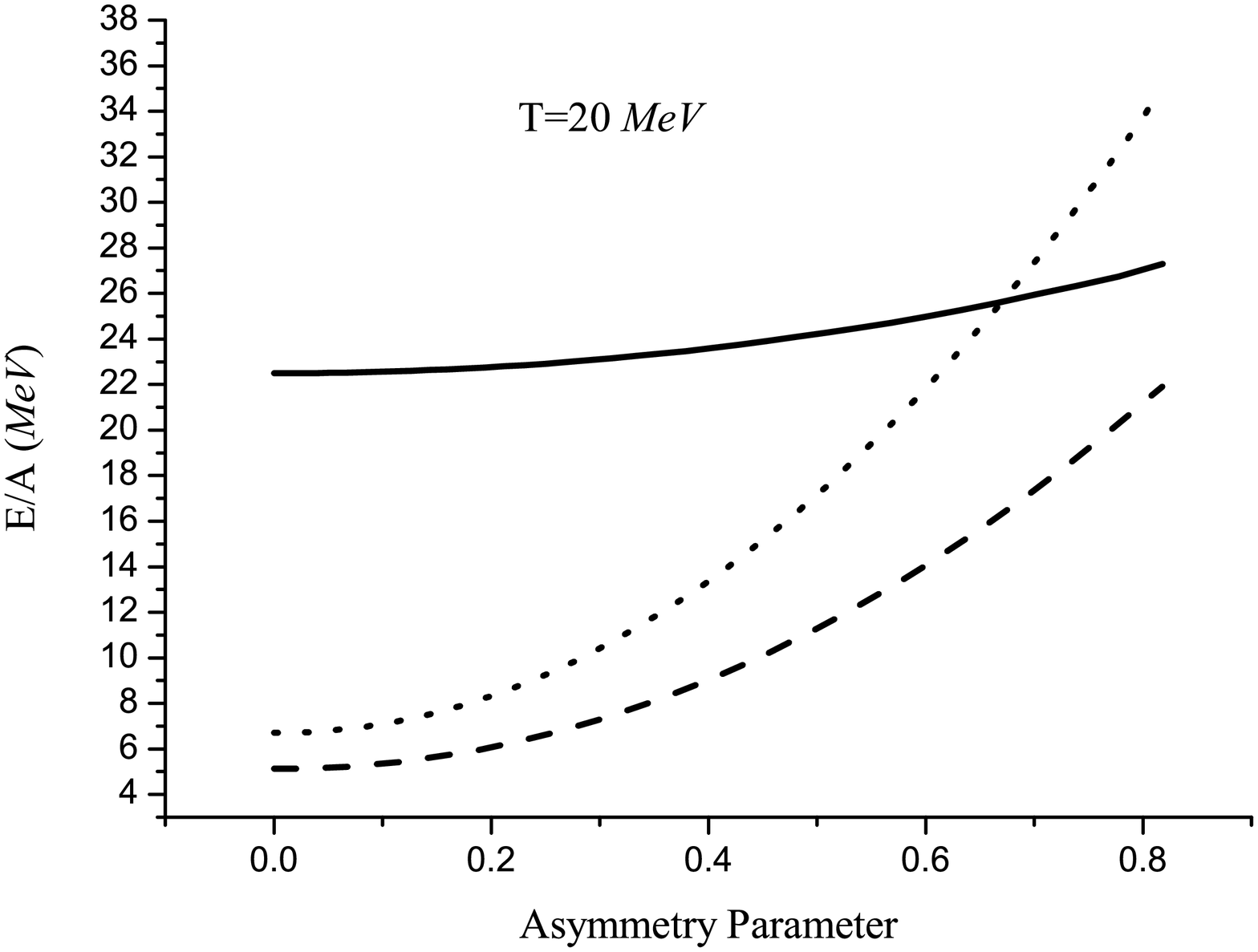}
\caption{The binding energy per nucleon of asymmetric nuclear matter versus asymmetry parameter ($\xi$) for densities $\rho = 0.05$ (full curve), $0.25$ (dashed curve) and $0.5\ fm^{-3} $ (dotted curve) at different temperatures ($T$).}
\label{fig1}
\end{figure}

We have calculated the binding energy of asymmetric nuclear matter using the method discussed in the previous section at various densities and temperatures. Our results for the binding energy versus asymmetry parameter ($\xi=\frac{N-Z}{N+Z}$) have been shown in Fig. \ref{fig1} for different densities and temperatures. For all densities and temperatures, it is seen that the binding energy increases as the asymmetry parameter increases.
We have found that for each value of temperature and asymmetry parameter, below a specific value of density, the binding energy decreases
by increasing density, while above this density, it increases as density increases.
This specific density is called the saturation density which depends on both temperature and asymmetry parameter.
For each density, by comparison of the energy curves, we observe that by increasing the temperature, the slope of energy curve versus asymmetry parameter decreases.
Fig. \ref{fig1} shows that for each temperature, the increasing rate of binding energy versus asymmetry parameter has different values for different densities.

For each value of density at a specific temperature, we can write our results for the binding energy versus asymmetry parameter ($\xi$) as a polynomial function in term of different powers of $\xi$.
By comparing the extracted polynomial function with the semi-empirical mass formula used for the binding energy of asymmetric
nuclear matter ($E=a_v+ a_{sym}\xi^2$),
the requirement of presence of other powers of $\xi$ can be analyzed, and it is determined whether
the semi-empirical mass formula would be broken or not.
We have found that the best function which is fitted with our results for the binding energy of asymmetric
nuclear matter
is a polynomial up to the fourth power of $\xi$,
\begin{equation}\label{poly}
E=a_0+a_1\xi+a_2\xi^2+a_3\xi^3+a_4\xi^4.
\end{equation}
The coefficients of various powers of $\xi$ ($a_0,\ a_1,\ a_2,\ a_3$ and $a_4$) versus density at different temperatures have been given
in Tables \ref{Tab1}, \ref{Tab2} and \ref{Tab3}.
By investigating the values of $a_0$ from these tables,
we observe that for each temperature, $a_{0}$ decreases by increasing the density up to a specific density, and then above this density, $a_{0}$ increases.
By comparing these tables,
we can find that for each density, with an increase in temperature, an increasing in $a_{0}$ is seen.
From Tables \ref{Tab1}, \ref{Tab2} and \ref{Tab3}, by investigating the coefficients $ a_1$, $a_3$ and $a_4$, we can conclude that
these coefficient are density and temperature dependent.
\begin{table}[h!t]%
\caption{$a_0$, $a_1$, $a_2$, $a_3$ and $a_4$ ($MeV$) versus density ($\rho$) for temperature $T=0\ MeV$.}
\centering
\begin{tabular}{|c|c|c|c|c|c|}
\hline
$\rho\ (fm^{-3})$& $a_0$& $a_1$& $a_2$& $a_3$& $a_4$\\
\hline
    0.05& 4.921& -0.05167 &12.01 &-0.117 &2.574  \\
    0.1& -10.08& -0.01954 & 17.64 & 1.548 & 0.4468 \\
    0.15&-13.76& -0.1445 & 23.76 & 0.1695 & 1.371  \\
    0.2& -16.3& -0.2257 & 28.82 & -0.6103 & 2.067 \\
    0.25& -17.85& -0.3245 & 33.57 & -1.974 & 3.028 \\
    0.3& -18.45& -0.4127 & 37.37 & -1.949 & 3.02 \\
    0.35& -18.19& -0.4084 & 40.68 & -1.887 & 3.007 \\
    0.4& -17.1& -0.5143 & 44.33 & -2.762 & 3.524 \\
    0.45& -15.23& -0.6283 & 47.94 & -3.658 & 3.932 \\
    0.5& -12.61& -0.6323 & 50.63 & -2.82 & 3.355 \\
    \hline
 \end{tabular}
\label{Tab1}
\end{table}
\begin{table}[h!t]%
\caption{$a_0$, $a_1$, $a_2$, $a_3$ and $a_4$ ($MeV$) versus density ($\rho$) for temperature $T=10\ MeV$.}
\centering
\begin{tabular}{|c|c|c|c|c|c|}
\hline
$\rho\ (fm^{-3})$& $a_0$& $a_1$& $a_2$& $a_3$& $a_4$\\
 \hline
    0.05& 6.73012 & -0.04794 & 8.94846 & 0.48205 & 0.75552 \\
    0.1& -0.02839 & -0.17483 & 15.43355 & -0.33542 & 0.77439 \\
    0.15& -4.86116 & -0.13702 & 20.64022 & 0.60392 & 0.2858  \\
    0.2& -8.2671 & -0.26539 & 26.04789 & -0.0697 & 0.79922  \\
    0.25& -10.48181 & -0.24034 & 30.19788 & 0.29485 & 0.71646 \\
    0.3& -11.63058 & -0.40422 & 34.81898 & -0.85014 & 1.4016 \\
    0.35& -11.79859 & -0.62335 & 39.13478 & -1.82609 & 1.88743 \\
    0.4& -11.07976 & -0.47165 & 41.84846 &-0.96744 & 1.5085 \\
    0.45& -9.50737 & -0.5509 & 45.37664 & -1.40054 & 1.66413 \\
    0.5& -7.1490 & -0.52195 & 47.92776 & 0.1053 & 0.63481 \\
    \hline
 \end{tabular}
\label{Tab2}
\end{table}
\begin{table}[h!t]%
\caption{$a_0$, $a_1$, $a_2$, $a_3$ and $a_4$ ($MeV$) versus density ($\rho$) for temperature $T=20\ MeV$.}
\centering
\begin{tabular}{|c|c|c|c|c|c|}
\hline
$\rho\ (fm^{-3})$& $a_0$& $a_1$& $a_2$& $a_3$& $a_4$\\
 \hline
    0.05& 22.50472 & -0.21782 & 8.02506 & -2.70781 & 2.36938 \\
    0.1& 16.14381 & -0.12781 & 11.66631 & 0.04014 & 0.33926 \\
    0.15& 11.29118 & -0.09781 & 15.88062 & 1.21584 & -0.27871  \\
    0.2& 7.65327 & -0.18837 & 20.78746 & 0.18392 & 0.49755 \\
    0.25& 5.12343 & -0.24998 & 24.86982 & 0.36818 & 0.36781 \\
    0.3& 3.61406 & -0.39247 & 29.43257 & -1.14565 & 1.30578 \\
    0.35& 3.07363 & -0.34095 & 32.36919 & 0.23839 & 0.38758 \\
    0.4& 3.45085 & -0.56422 & 36.75817 & -1.36493 & 1.19642 \\
    0.45& 4.68426 & -0.55863 & 39.84991 & -0.73474 & 0.63449 \\
    0.5& 6.71498 & -0.48627 & 42.39283 & 1.09249 & -0.69099 \\
    \hline
 \end{tabular}
\label{Tab3}
\end{table}
The given values of $a_2$ in these tables indicate that for each temperature, the value of $a_{2}$ increases by increasing the density. We have found that at  high densities, the increasing rate of $a_{2}$ is lower than that at low densities.
A comparison between the values of $a_{2}$ in different temperatures shows that for all densities, the value of $a_{2}$ decreases as the temperature increases. Here, we can conclude that the value of $a_{2}$ depends on both density and temperature.

Now, by more investigating Tables \ref{Tab1}, \ref{Tab2} and \ref{Tab3}, we see that for all densities and temperatures,
with respect to the other coefficients, the coefficients
$ a_0$ and $ a_2$ have the main contributions in the binding energy.
However, we test the the effects of $a_{1}$, $a_{3}$ and $a_{4}$ relative to $a_{2}$ as follows.
By obtaining the magnitude of ratio $\frac{a_1}{a_2}$, it is clear
that at low densities, this ratio is of order $10^{-3}$, therefore the effect of $a_{1}$ in the binding energy of asymmetric nuclear matter is negligible.
We have found that at high densities, this ratio is of order $10^{-1}$, therefore it is better to consider
the effect of $a_{1}$ for the accurate calculations.
By comparing the magnitude of $a_{3}$ and $a_{4}$, it is seen that for all temperatures and densities,
the magnitude of $a_{4}$ is always greater than that of $a_3$.
Our results show that the highest order of magnitude of ratio $\frac{a_3}{a_2}$ reaches about $10^{-2}$.
For the magnitude of  $\frac{a_4}{a_2}$, that is of order $10^{-1}$.
The above results indicate that the dominant term in the asymmetry energy of nuclear matter is related to $a_{2}$, and the other terms which are related to $a_{1}$, $a_{3}$ and $a_{4}$ are small with respect to that of $a_{2}$. However, in order to do more accurate calculation, these terms could be considered.

Here, for the binding energy of asymmetric nuclear matter, we rewrite Eq. (\ref{poly}) as the following relation
\begin{equation}
E= a_v+E_{asym},
\end{equation}
where $a_v=a_{0}$ is the volume energy and $E_{asym}$ is the asymmetry energy
(for the symmetric nuclear matter in which $N=Z$, $E_{asym}$ is zero),
\begin{equation}
E_{asym}= E_{asym}^{(1)}+ E_{asym}^{(2)}+E_{asym}^{(3)}+E_{asym}^{(4)}.
\end{equation}
In fact the contribution of asymmetry energy is considered as four terms.
Now,  at a specific density and temperature, we compare the effects of different terms of $E_{asym}$ with respect to  $E_{asym}^{(2)}$ for
the low ($\xi=0.1$) and high ($\xi=0.9$) values of
asymmetry parameter. Our results indicate that for the low values
of asymmetry parameter, the magnitude of ratio $\frac{E_{asym}^{(1)}}{E_{asym}^{(2)}}$ is significant,
especially at higher densities, therefore, for this case of asymmetric nuclear matter,
 it is better to consider the effects of $E_{asym}^{(1)}$. However, by increasing
the asymmetry parameter, the effect of this term decreases.
The magnitude of ratio $\frac{E_{asym}^{(3)}}{E_{asym}^{(2)}}$ and $\frac{E_{asym}^{(4)}}{E_{asym}^{(2)}}$ at
low value of asymmetry parameter, is very small. The magnitude of these ratios at high values of asymmetry parameter becomes more important,
but they are yet small.

\section{Summary and Conclusions}
We have considered the asymmetric nuclear matter including $N$ neutrons and $Z$ protons which interact through the nuclear force. For this system, the binding energy has been computed by a fully microscopic variational many-body technique based on the cluster expansion of the energy. In our formalism, we have presented the $AV_{18}$ two-nucleon potential.
We have also calculated the asymmetry energy of nuclear matter. It is found that the asymmetry energy is a function of density and temperature.
For different densities and temperatures, our results for the binding energy versus asymmetry parameter have been fitted as a polynomial function.
It was a polynomial of forth power of asymmetry parameter.
We have seen that all the coefficients of this polynomial are density and temperature dependent.
By comparing our polynomial function with the semi-empirical mass formula used for the binding energy of asymmetric nuclear matter,
the requirement of presence of other powers of asymmetry parameter has been analyzed, and the validity of semi-empirical mass formula has been investigated.
According to our results, the square term in the asymmetry energy of nuclear matter is the dominant term, and the other terms compared to this term are small.
Therefore, we can ignore the effects of non-square terms.
Our study shows that the asymmetry energy of asymmetric nuclear matter has a relatively good agreement with the semi-empirical mass formula.
However, according to our calculations, to increase the accuracy of calculations, it is better to include the other terms in the asymmetry energy.
For example, the first order term at low values of asymmetry parameter is considerable, especially at low densities.

\acknowledgements{
This work has been supported financially by the Center for Excellence in
Astronomy and Astrophysics (CEAA-RIAAM). We wish to thank the Shiraz University
Research Council.}

\newpage

\end{document}